\documentclass[journal]{IEEEtran}
\usepackage{amssymb}
\usepackage{graphicx}
\newtheorem{theorem}{Theorem}
\newtheorem{example}[theorem]{Example}
\begin{document}
\title{LiteMat: a scalable, cost-efficient inference encoding scheme for large RDF graphs}

\author{\IEEEauthorblockN{Olivier Cur\'e,}
\and
\IEEEauthorblockN{Hubert Naacke,}
\and
\IEEEauthorblockN{Tendry Randriamalala,}
\and
\IEEEauthorblockN{Bernd Amann}\\
\IEEEauthorblockA{LIP6 CNRS UMR 7606\\
Sorbonne Universit\'es, UPMC Univ Paris 06, F-75005, Paris, France\\
Email: \{firstname.lastname\}@lip6.fr}}

\maketitle

\begin{abstract}
The number of linked data sources and the size of the linked open data graph keep growing every day. As a consequence, semantic RDF services are more and more confronted with various "big data" problems. 
Query processing in the presence of inferences is one them. For instance, to complete the answer set of SPARQL queries, RDF database systems evaluate semantic RDFS relationships (subPropertyOf, subClassOf) through time-consuming query rewriting algorithms or space-consuming data materialization solutions.
To reduce the memory footprint and ease the exchange of large datasets, these systems generally apply a dictionary approach for compressing triple data sizes by replacing resource identifiers (IRIs), blank nodes and literals with integer values. 
In this article, we present a structured resource identification scheme using a clever encoding of concepts and property hierarchies for efficiently evaluating the main common RDFS entailment rules while minimizing  triple materialization and query rewriting. We will show how this encoding can be computed by a scalable parallel algorithm and directly be implemented over the Apache Spark framework. The efficiency of our encoding scheme is emphasized by an evaluation conducted over both synthetic and real world datasets.

\end{abstract}

\section{Introduction}
The Resource Description Framework (RDF) data model is the W3C standard for representing metadata in the Web of Data and the Semantic Web. This format is used to form very large graphs, such as those found in Linked Data sources and Linked Open Data (LOD), which range in hundreds of millions to billions of RDF triples. With such workloads, RDF database systems are now facing "big data" problems. 
In this work, we focus on issues related to SPARQL query processing over large RDF data with rich semantics. 
One aspect that differentiates RDF data, and Knowledge bases (KB) in general, from standard relational and NoSQL systems, 
e.g., graph databases such as Neo4J and Titan, is the ability to reason over the represented information using associated knowledge. Such knowledge, contained in ontologies, is generally expressed using RDF Schema (RDFS) or Web Ontology Language (OWL) W3C recommendations.
From a query processing point of view, to achieve completeness of result sets, RDF query processors have to integrate information that is inferred using these ontologies. 
We can distinguish two main approaches to support this kind of inference. The first approach
consists of materializing all derivable triples in the RDF store before evaluating queries. The second approach consists of rewriting each submitted query into an extended query including semantic relationships from the ontologies. Both methods have advantages and drawbacks. 

On one hand, materialization implies a possibly long loading time due to running reasoning services during a data pre-processing phase. This generally drastically increases the size of the stored data and imposes specific dynamic inference strategies when data is updated.  The advantage is that it ensures good query performance. 

On the other hand, query rewriting avoids costly data pre-processing, storage extension and complex update strategies but induces slow query response times since all the reasoning tasks are part of a complex query pre-processing step. 

In the worst case the computational complexity associated to data materialization and query rewriting processing are exponential in the size of the original data and query respectively. 
Some existing techniques \cite{DBLP:conf/kr/Rodriguez-MuroC12} propose polynomial solutions in specific situations. 
We now provide an example to make these two approaches more concrete.

\begin{example}We consider a simple example based on the following extract of the LUBM ontology \cite{DBLP:journals/ws/GuoPH05}. The concept hierarchy  (TBox, short for Terminological Box) is limited to the following axioms:

(1) $Professor \sqsubseteq FacultyMember $

(2) $\exists teaches.\top \sqsubseteq FacultyMember$

where $\sqsubseteq$ denotes the subsumption relationship, i.e., a $Professor$ is a subconcept of $FacultyMember$. The domain of the property $teaches$  is the concept  $FacultyMember$. The ABox (short for Assertional Box), is a set of facts and consists of the following RDF triples:

(3) $bernd~ type~ Professor.$
(4) $hubert~ teaches~ course1.$

The SPARQL query aiming to retrieve all $FacultyMember$ instances is expressed as follows:

$SELECT~ ?x~ WHERE~ \{?x~ type~ FacultyMember\} $

The complete and correct answer set with respect to the TBox contains both $bernd$ and $hubert$ but without any inference, the query would return an empty result set since none of the instances, i.e., $bernd$, $hubert$ or $course1$, are explicitly typed with the concept $FacultyMember$. 
In the materialization approach the ABox would be extended to contain the following triples:

(5) $bernd~ type~ FacultyMember.$ 

(6) $hubert~ type~ FacultyMember.$

where (5) is deduced from (1) and (3) while (6) is derived from (2) and (4).

A complete and sound answer set can be retrieved with the following reformulated query: 

$SELECT~  ?x~  WHERE~  \{\{ ?x~  type~  FacultyMember.\}$\\
$UNION~  \{?x~  teaches~  ?y.\} UNION~ \{?x~  type~ Professor.\}\}$
\end{example}

In this article, we present the LiteMat data encoding and SPARQL query evaluation approach which aims at finding an efficient trade-off between materialization and query rewriting that provides complete answer sets considering the RDFS entailment regime. The LiteMat approach builds on a semantic-aware RDF data encoding which allows to reduce the amount of materialized information with minimal query reformulation and processing cost. The approach is based on a clever encoding of the ontology elements of the KB and only requires some functions to check whether a variable binding belongs to a numerical interval.
We will demonstrate that LiteMat can reduce the number of triples of an original dataset.

%
%
Most RDF stores, e.g. \cite{DBLP:journals/vldb/NeumannW10}, adopt an encoding approach to compress RDF triples. 
This is motivated by the fact that components of RDF triples, i.e., subjects, predicates and objects, mainly correspond to Internationalized Resource Identifiers (IRI) or literals. They correspond to long strings of characters and the number of their occurrence can be quite important in real world cases. 
Thus replacing them with integer values permits to obtain a much more compact representation of a set of triples. Blank nodes are encoded using a similar approach.

Most RDF systems use dictionary-based encoding for long IRI resource identifiers but do not consider semantic aspects when attributing integer values to RDF entities (see \cite{cure:hal-01083133} for more details on this topic). 
That is, they do not consider the ontology at this processing stage. 
As a result, no distinctions are made between an instance or concept/predicate IRI. 
Our solution uses that distinction and is based on a clever encoding of the TBox which is implied in the encoding of the ABox. 

Given the size of currently available RDF datasets, this encoding step can result in a performance bottleneck. To speed up this processing, a parallel computation may be necessary. Once the encoding is performed, two operations are needed to handle an encoded dataset: $locate$ and $extract$. The $locate$ operation takes as input a string (IRI, blank node identifier or literal) and provides the corresponding id. 
The $extract$ operation is provided with an id and returns a string value. 
These operations generally handle key/value pairs.

The main contributions of this paper are to propose an encoding algorithm for the TBox, to present a generic scalable algorithm for the encoding of ABoxes, to provide a lite materialization strategy together with its query processor supporting SPARQL's RDFS entailment regime. 
Finally, to tackle very large graphs, we evaluate our implementation over the Apache Spark framework using synthetic and real world use cases.

\section{Background knowledge}
\subsection{RDF, SPARQL, RDFS entailment regime}
RDF is a schema-free data model that supports the description of data on the Web. 
It is usually considered as the cornerstone of the Semantic Web and the Web of Data.
Assuming disjoint infinite sets U (RDF IRI references), B (blank nodes) and L (literals), a triple (s,p,o) $\in$ (U $\cup$ B) $\times$ U $\times$ (U $\cup$ B $\cup$ L) is called an RDF triple with s, p and o 
respectively being the subject, predicate and object. 
We now also assume that V is an infinite set of variables and that it is disjoint with U, B and L. 
We can recursively define a SPARQL\footnote{http://www.w3.org/TR/rdf-sparql-query/} triple pattern as follows: 
(i) a triple $tp \in$ (U $\cup$ V) $\times$ (U $\cup$ V) $\times$ (U $\cup$ V $\cup$ L) is a SPARQL triple pattern, (ii) if $tp_1$ and $tp_2$ are triple patterns, then ($tp_1 . tp_2$) represents a group of triple patterns 
that must all match, ($tp_1$ \texttt{OPTIONAL} $tp_2$) where $tp_2$ is a set of patterns that may extend the solution induced by $tp_1$, and ($tp_1$ \texttt{UNION} $tp_2$), denoting pattern alternatives, are 
triple patterns and (iii) if $tp$ is a triple pattern and C is a built-in condition then the expression ($tp$  \texttt{FILTER} C) is a triple pattern 
that enables to restrict the solutions of a triple pattern match according to the expression C.
The SPARQL syntax follows the select-from-where approach of SQL queries. The \texttt{SELECT} clause specifies the variables appearing in the result set of the query.

The SPARQL 1.1 W3C recommendation specifies various entailment regimes which define the evaluation of SPARQL
 triple patterns by means of subgraph matching. 
In this work we are interested in the RDFS entailment regime. 
It is based on the set of reasoning rules of the RDFS language\footnote{http://www.w3.org/TR/rdf11-mt/}. 
We concentrate on property and class subsumption, domain and range of properties. 
They have been selected due to their support of the most frequent RDFS inferences.

\subsection{Apache Spark}
\label{sec:spark}
Apache Spark \cite{DBLP:conf/hotcloud/ZahariaCFSS10} is a cluster computing framework based on a shared nothing architecture.
Just like Apache Hadoop, Spark enables parallel computations on unreliable machines and automatically handles locality-aware scheduling, fault tolerance and load balancing. While both systems are based on a data flow computation model, Spark is more efficient than Hadoop for applications requiring to reuse working datasets across multiple parallel operations.
This efficiency is due to Spark's Resilient Distributed Dataset (RDD) \cite{DBLP:conf/nsdi/ZahariaCDDMMFSS12}, a distributed, lineage supported fault tolerant memory abstraction that enables one to perform in-memory computations (when Hadoop is mainly disk-based). The Spark API also eases the programming task by integrating functions which are not natively supported in Hadoop, $e.g.,$ join, filter.  The design and implementation of Spark started at UC Berkeley's AMPLab and at the time of writing this paper, it is considered to be the Apache project with the most committers.

\section{Knowledge Base encoding}
\label{sec:encoding}
The encoding principle of our approach is similar to the ones encountered in most RDF Stores. 
A first phase creates a dictionary corresponding to a bijective function mapping long terms (IRI, blank node identifier, literal) to short identifiers (integer). 
The dictionary is then managed using the $locate$ and $extract$ operations previously introduced.
 
The originality of our approach is that we distinguish the TBox encoding from the ABox one. 
For the former, we propose a semantic-aware encoding which supports the lite materialization approach. 
While this phase is performed on a single machine of the cluster, the ABox encoding is computed in parallel using Spark. The values we assign for entities of the TBox and ABox can overlap, i.e., the same value can be given to an individual, a property and a concept. This does not cause any problems in our system due to context awareness at query processing time. For instance, we know that each identifier at the second position of an RDF triple is necessarily a property and that objects associated to an \texttt{rdf:type} property are necessarily concepts. All other identifiers in the ABox correspond to individuals.

\subsection{TBox encoding}
\subsubsection{Principle}
Our TBox encoding principle applies to both the concept and property hierarchies, which we denote as entities in the following. The idea is to assign an integer value to each entity such that for a sub-entity B of an entity A ($B \sqsubseteq A$), B's assigned identifier ($idB$) is comprised in an interval that can be easily and automatically compute from A's identifier (idA). That is $idB \in~]idA, idA +\epsilon[$ where $\epsilon$ depends on the number of direct sub-entities of the entity A.
This encoding scheme applies recursively to the complete entity hierarchy.

This encoding scheme covers both the \texttt{rdfs:subClassOf} and \texttt{rdfs:subPropertyOf} associated entailments. To support inference rules related to \texttt{rdfs:domain} and \texttt{rdfs:range}, we define two additional data structures, one for each RDFS property, consisting of a key/value pair. The key corresponds to the property identifier and the value contains the set of this property's domain (resp. range) concept identifiers.

\subsubsection{Implementation}
In order to grasp accurate entity hierarchies, we use an OWL reasoner to infer concept classifications. This implies that we are representing hierarchies that are not limited to the interpretation of the RDFS language. Nevertheless, based on this representation, we only consider the RDFS entailment regime, e.g., currently our system is not able to handle inference rules related to transitive properties.

HermiT \cite{msh09hypertableau}, the OWL reasoner we are using for the encoding, does not support a distributed computing approach. We thus generate the encoding of a TBox on a single machine.
The algorithm represents entity identifiers as vectors of bits and consists of two phases. 
In the first phase, a top-down navigation of the inferred entity hierarchy, e.g., for concepts, is adopted.
Intuitively, we start from the \texttt{owl:Thing} concept down to all subconcepts. 
Given an entity A, we first compute the number N of direct sub-entities. 
These sub-entities will be encoded over $\lceil log_2(N+1) \rceil$ bits. 
This is performed recursively until all entities in the TBox are  assigned to an identifier. 
It is guaranteed at the end of this first phase that, for 2 entities A and B with $B \sqsubseteq A$, the prefix of $idB$ matches with the encoding $idA$. 

At the end of this first step, all entities have been given a prefix value. 
Table \ref{lubmEncoding} presents an extract of the encoding of the LUBM ontology. For instance, in LUBM, the \texttt{Schedule} and \texttt{AssociateProfessor} are respectively given identifiers '0001' and '01001110010011'. 
Given that \texttt{owl:Thing} is always assigned the value 0, we can see that its 5 direct subconcepts (\texttt{Schedule}, \texttt{Organization}, \texttt{Publication}, \texttt{Person} and \texttt{Work}) are encoded over 3 bits. Moreover, \texttt{AssociateProfessor} shares the prefix '0100' with its indirect \texttt{Person} ancestor.  In order to support our interval operations over identifiers, we need to represent these bit vectors over the same number of bits. The total number of bits corresponds to the encoding of the longest entity identifer, i.e., 14 bits for the \texttt{AssociateProfessor} or any of its siblings. Note that encodings for DBPedia and Wikidata's concept hierarchies required respectively 27 and 102 bits, justifying the use of vector bits for the latter.
This second step handles the identifier completion by appending '0' on rightmost bits. 

\begin{table}
\begin{center}
\begin{tabular}{|l|c|}
\hline
Encoding & Concept label\\
\hline
0 000 0000000000 & Thing\\
\hline
0 001 0000000000 & Schedule\\
\hline
0 010 0000000000 & Organization\\
\hline
0 011 0000000000 & Publication\\
\hline
0 100 0000000000 & Person\\
\hline
0 100 0100000000  & TeachingAssistant \\
\hline
0 100 1000000000 & Student\\
\hline
...& \\
\hline
0 100 1110010011 & AssociateProfessor\\
\hline
...&\\
\hline
0 101 0000000000 & Work\\
\hline
\end{tabular}

\caption{Extract of the LUBM concept encoding}
\label{lubmEncoding}
\end{center}
\end{table}

\subsection{ABox encoding}

\subsubsection{Principle}
The encoding of the ABox is more straightforward than the TBox encoding since it does not require an external software component, i.e., a reasoner.
The idea is to provide a unique integer identifier to all ABox entries that have not been assigned an identifier during the TBox encoding. Since we are not giving any 'meaning' to these values, the encoding can be performed in parallel. This presents a particular advantage due to the size of the ABox which can be orders of magnitude larger than the TBox.

\subsubsection{Implementation}
For the ABOX encoding we will benefit from the Spark framework by storing most of the processed datasets, i.e., the portion corresponding to the ABox, in main-memory and by taking advantage of a rich API of operators including join, union, duplicate elimination.
In a first stage, the dataset is uploaded from the Hadoop Distributed File System (HDFS) and partitioned across the cluster. 
The system identifies the  subject/object and possible property terms which are not covered by the TBox encoding. 
Then each distinct entry is assigned an identifier in a distributed manner. Intuitively, we create an array that contains the number of subjects and objects contained in each partition. These values are summed up so that each partition can be associated with a disjoint identifier interval. Then each partition provides identifiers to all its members ranging over the partition's identifier interval.
This set of subject/object (resp. properties) is then unioned with the concept (resp. property) encoding emanating from the TBox.

Given these two maps, the original string-based dataset is encoded via some join operations whose execution can be optimized by replicating the maps over the cluster nodes, i.e., storing the maps in the main-memory of all executors of the cluster and thus preventing network communications. In our evaluation, the property map is broadcasted while for some cases, e.g., Wikidata (Table \ref{ontologies}), broadcasting the concept map would be counterproductive due to its large memory footprint.

\section{Lite Materialization}
\label{sec:liteMat}
Our TBox encoding scheme provides the nice property of retaining only the most specific concept in the hierarchy while still being able to perform some subsumption inferences. In this context, our LiteMat approach aims to minimize the amount of types assigned to each instance in the dataset.
This may be useful since some real world datasets are known to have several types for a given instance, e.g., an average of 8 different types are provided on DBPedia individuals.
The information stored in our TBox encoding data structures enables to perform this materialization.

The parallel algorithm proceeds in 2 steps.
In the first step, for each individual $I$, the system gathers all explicit and implicit types where the former correspond to types already stored in the original datasets and the latter correspond to the ones that can be derived from the domain and range of RDFS properties involved in $I$'s assertions. This set of types is henceforth denoted as candidate concepts.
For instance, in Example 1, the domain of \texttt{teaches} enables to attribute the \texttt{FacultyMember} type to the $hubert$ instance.

\begin{figure}
\centering
\includegraphics[scale=0.5]{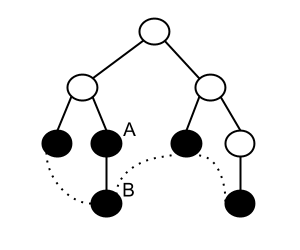}
\caption{Candidate concepts in their hierarchy}
\label{liteMatGraph}
\end{figure}

In Figure \ref{liteMatGraph}, we represent a set of such concepts in the context of their subsumption hierarchies. 
Each node represents a TBox concept with the black ones being the candidate concepts and the white ones being their super concepts.
The plain lines represent subsumption hierarchies with the lower node always being the subconcept.
To minimize the number of types for a given instance, the system needs to retain only the Most Specific Concepts (MSC) in each subsumption branch of candidate concepts.
This corresponds to identifying the nodes going through the dashed line in Figure \ref{liteMatGraph}.
Note that during this identification process, we are not considering the complete concept hierarchy but only the candidate concepts and the transitive closure 
of their super concepts.

Our system performs this operation in one pass using our TBox encoding data structures. 
Intuitively, set of candidate concepts of an instance $I$ is ordered in descending order on the values of their identifiers. 
This is motivated by our encoding scheme which ensures that a subconcept has a greater value than its super concept.
The first concept in that list is stored in the MSC set and subsequent candidates are checked for insertion in the MSC set.
A candidate is inserted in the MSC set if none of the concepts in the MSC set belongs to its  subsumption interval.

A concept's 'subsumption interval' is computed using the concept id and 3 meta data corresponding to its total encoding length (codeLength),  the position at which this concept encoding starts in the bit vector (start) and the length of its encoding (localLength). Given these information, the following $bound$ function returns the upper bound of a concept.

\begin {verbatim}
def bound(id : String, start: Int, 
          localLength: Int, codeLength: Int)
: (String) = {
  val shift = codeLength - 
              (start+localLength)
  val prefix = id >> shift
  val upperBound = (prefix+1) << shift 
  return upperBound
}
\end{verbatim}

In Figure \ref{liteMatGraph}, the concept A is not retained in the MSC set since B is more specific. 
Consider the following setting: A and B have respective identifiers: 20 (00010100) and 22 (00010110), 
the encoding length is 8 bits, the encoding of A starts at bit 3 and is encoded using 3 bits. 
Then the subsumption interval of A is $[20,24[$. 
This justifies that A is not in the final MSC since B with an identifier of 22 belongs to that interval.

We finally would like to highlight that the lite materialization may also have the positive effect of diminishing the original dataset by removing some unnecessary concepts (one of our experimentation highlights such a situation on the Wikidata real world dataset).

\section{Query processing}
\label{sec:query}
This section details our evaluation method of conjunctive SPARQL queries using the encoding scheme described in Section~\ref{sec:encoding}.
This method guarantees the completeness of the query result. It considers the RDFS class and property hierarchies, i.e., considers all triples that match a  graph pattern according to these hierarchies. 
More precisely, the triple matching procedure takes into account the following semantic inference rules:
\begin{itemize}
\item a triple $(s, p, o)$ matches the triple pattern  $(?x, prop, ?y)$ if $p$ is a sub-property of $prop$ (by definition $p$ is a subproperty of itself).
\item a triple \textit{(a, rdf:type, b)} matches the triple pattern \textit{(x, rdf:type, c)} if $b$ is a subtype of $c$ (by definition, $b$ is a subconcept of itself).
\end{itemize}

These rules can be efficiently evaluated by the semantic-aware encoding described in Sec.~\ref{sec:encoding}. 
It allows the query processor to compare straightforwardly the values of $p$ and $prop$ (and respectively the values of $b$ and $c$), based on the encoded values and the previously defined $bound$ function (Sec.~\ref{sec:liteMat}): 
\begin{equation}
p~is~a~subpropertyof~prop \Leftrightarrow  prop \leq p < bound(prop)
\end{equation}
\begin{equation}
b~is~a~subtype~of~c \Leftrightarrow c \leq b < bound(c)
\end{equation}

Note that we implement subproperty and subtype matching as a single comparison operation without testing every existing subproperty and subtype. This obviously saves computation cost. 

Our query plan generation method targets the Apache Spark distributed and parallel computing platform for processing queries. As explained in Sec.~\ref{sec:spark}, Spark provides efficient methods to manipulate distributed datasets in parallel.
We use the following Spark operations applied to datasets: 
\begin{itemize}
\item dataset.filter(predicate): select the subset of the dataset that satisfies a predicate. 
\item dataset.map(function): apply a function on each element of the dataset.
\item dataset.join(dataset): join two datasets using an equality predicate. It requires that the structure of the elements in each dataset is a pair. For instance (a,b) join (c,d) produces a result if a=c.
\end{itemize}
It has been showed in \cite{spark:sql} that these operations are sufficient to express any conjunctive query. 
With this in mind, we describe the translation of conjunctive SPARQL query into an algebraic expression.


\textbf{Locate step}. In a typical use case, a SPARQL engine calls the $locate$ function to transform the constants of a triple pattern into numerical values.
In our case, we translate every term (property, type and literal) of the query into its corresponding encoded value. 

\begin{example}
The execution of the $locate$ function over Query \#4 in \ref{appendix:query} would yield the following triple patterns:
\begin{verbatim}
?x 0 1044643840. ?y 0 335544320.
?x 1145044992 ?y.
\end{verbatim}

\end{example}

\textbf{Matching step}. We translate each query triple, into a filter operation that implements the matching. For clarity, we detail below the resulting algebraic expression written in (simplified) Scala programming language. The name $triples$ denotes the dataset.\\
(?x, prop, ?y) is translated into  
\begin{verbatim}
triples.filter( (s,p,o) => 
   prop <= p && p < bound(prop) )
\end{verbatim}
(?x, rdf:type, c) is translated into 
\begin{verbatim}
triples.filter( (s,p,o) => 
   p == type && c <= o && o < bound(c) )
\end{verbatim}

\textbf{Conjunction step}. 
We translate the conjunction of two triple patterns into a join expression. We denote t1 (resp. t2) the expression resulting from the translation of the first (resp. the second) triple pattern\\
(?x, $prop_1$, ?y) . (?y, $prop_2$, ?z)  is translated into 
\begin{verbatim}
t1.map( (s,p,o) => (o, (s,p)).
   join(t2.map( (s,p,o) => (s, (p,o)))
\end{verbatim}

The map function allows to specify on which variable the equality join must apply. Our translation method supports all the join cases: subject-subject, subject-object, and object-object.
We handle several conjunctions by successively applying that translation on each pattern. Note that the finding an optimal join ordering for triple patterns is beyond the scope of this article.






\textbf{Extract step}. 
The last query processing step consists of calling the $extract$ function to translate the result set (so far expressed with numerical values) into an end-user understandable form, i.e., IRIs, literals and possibly blank node identifiers.


\section{Evaluation}
\subsection{Computational environment}
The evaluation was conducted on a cluster consisting of 15 DELL PowerEdge R410 running a Debian GNU/Linux distribution with a 3.16.0-4-amd64 kernel version. Each machine has 64GB of DDR3 RAM, two Intel Xeon E5645 processors. Each processor is constituted of 6 cores running at 2.40GHz and allowing to run two threads in parallel (hyper threading). Hence, the number of virtual cores amounts to 24.
Concerning storage, each machine is equipped with a 900GB 7200rpm SATA disk. The machines are connected via a 1GB/s Ethernet network adapter.
We used Spark version 1.4.1 and implemented all experiments in Scala, using version 2.10.4. 
The Spark configuration of our evaluation enables to run our prototype on a subset of the cluster corresponding to 300 cores and 50GB of RAM per machine.

\subsection{Datasets and queries}
We have selected 2 synthetic and 2 real world knowledge bases. The synthetic datasets correspond to instances of the LeHigh University Benchmark, a  well-established Semantic Web set  of tools composed of an ontology, a set of queries and a data generator. The knowledge bases have respectively been configured with 1000 and 10000 universities, resp. denoted LUBM1K and LUBM10K. The real world datasets correspond to  open source DBPedia and Wikidata RDF dumps. 

\begin{table*}
\begin{center}
\begin{tabular}{|c|c|r|c|r|r|c|}
\hline
Ontology & Size & \#Concepts & \#Prop. (Obj.+Datatype)  & \#Domain axioms & \#Range axioms & Total time in sec.\\
\hline
LUBM & 15KB & 43 & 32 (25+7) & 21 & 18 & 0.7 \\
\hline
DBPedia & 2.3MB & 814 & 3,035 (1,310+1,725) & 1,076 & 997 & 3.7\\
\hline
Wikidata & 78MB& 213,958 & 353 (255+98) & 0 & 0 & 122\\
\hline
\end{tabular}
\caption{Main features and encoding time of the evaluated ontologies }\label{ontologies}
\end{center}
\end{table*}

\begin{table*}
\begin{center}
\begin{tabular}{|c|r|r|r|r|r|}
\hline
Dataset & \#Triples \(* 10^6\) & SAE time (sec.) & SAE throughput (triples/sec.) & OBE time (sec.) & OBE throughput (triples/sec.)\\
\hline
LUBM1K & 133.5 &  280.5 & 476,199 & 113.6 & 1,157,486\\
\hline
LUBM10K & 1,334.7 &  3,746.5 & 356,248 & 1,755.8 & 759,378\\
\hline
DBPedia & 79.1 &  282.2 &  280 943  & 128 & 617,050  \\
\hline
Wikidata & 242.1 &  1,334.8 & 181,394 & 477.2 & 507,386\\
\hline
\end{tabular}
\caption{Comparison of the Standard ABox Encoding (SAE) with the Ontology-Based Encoding (OBE)}\label{encoding}
\end{center}
\end{table*}

In table~\ref{ontologies}, we present the main characteristics of the associated ontologies. Concerning DBPedia and Wikidata, the ontologies respectively correspond to DBPedia\_2014\footnote{http://oldwiki.dbpedia.org/Downloads2014} and the union of wikidata-taxonomy and wikidata-properties\footnote{http://tools.wmflabs.org/wikidata-exports/rdf/exports/20140526/}. We can observe that they differ in terms of their size (from 15KB to 78MB), number of concepts (from 43 to over 210K), properties (from around 30 to over 3K) and number of axioms related to property domain and range (from none to around 2K).
The main characteristics of the ABoxes are reported in the first two columns of Table~\ref{encoding}.

For the query processing evaluation, we choose a subset of the LUBM benchmark queries  that allow us to highlight the benefits of our LiteMat approach.

\subsection{Results}
In this section, we compare the encoding and query performance of our solution with a baseline solution based on a full materialization using a standard TBOX-unaware dictionary encoding.
\subsubsection{Encoding efficiency}
In order to evaluate LiteMat's encoding strategy, we present the performances of both our TBox and ABox encoding.
To evaluate the ABox encoding, we compare the standard ABox-only encoding (denoted SAE) which ignores any TBox knowledge, versus our ontology-based encoding solution (denoted OBE). SAE is completely computed in parallel over Spark. OBE is composed of an ontology encoding (implemented in Java and using the HermiT reasoner and the OWLAPI) and an ABox encoding (which runs over Spark in a manner quite similar to SAE) steps.

The rightmost column of Table~\ref{ontologies} presents the duration of the ontology encoding, i.e., processing the concept and property hierarchies as well as the structures for the domain and range RDFS properties, for the three ontologies under test. 
We report relatively fast execution times for all ontologies (up to 122 seconds for the largest ontology containing more than 200K concepts and over 300 properties). 

We also report the encoding duration and throughput (\textit{i.e.}, the number of RDF triple statements encoded per second) of the two encoding solutions in Table~\ref{encoding}.
Note that the reported times for OBE includes the TBox encoding time (Table~\ref{ontologies}). 
First we observe that both OBE and SAE solutions yield high throughput compared to the state-of-the-art solution reported in \cite{DBLP:journals/concurrency/UrbaniMDSB13}, thanks to our highly parallel implementation of the two solutions over the Spark platform.

Second, we observe that the OBE is up to 2.8 times faster than the SAE. 
In fact, for SAE, all individuals, concepts and properties have to be encoded. While for OBE, concepts and properties have already been encoded and are just uploaded for ABox encoding. The duration difference between the two approaches is explained by the high number of \textit{rdf:type} triples in most RDF datasets.




Moreover, for OBE, the size of the ontology unsurprisingly impacts the duration: although LUBM1K is 1.6 times larger than the DBPedia dataset, its encoding time is 12\% shorter.

Finally, we observe rather promising scalability of our OBE solution for growing datasets: the throughput is only decreasing by 35\% when the LUBM dataset size is ten times larger (ranging from 133M to 1.3B triples).


\subsubsection{Evaluation of materialization strategies}
In this section, we compare our LiteMat materialization approach with a standard full materialization solution.
The results are respectively reported in Tables~\ref{liteMat} and~\ref{fullMat}. 
First, for small dataset the duration is about the same, we observe a difference for larger KBs where the full materialization is much longer (between 1.5 and 2 times longer). 
Second, for LUBM, the dataset size remains almost the same. Indeed, the number of extra triples added during domain/range inference is slightly the same as the number of triples removed when retaining only the most-specific concepts.
%
Third, Considering the full materialization, unsurprisingly, the relative increase of the dataset size is ranging
from 13\% for DBPedia and up to almost 58\% for Wikidata.
This increase in size is directly related to the depth of ontology entity hierarchies, e.g., the DBPedia and Wikidata have several concept branches of depth larger than 6.

\begin{table*}
\begin{center}
\begin{tabular}{|c|r|r|r|r|}
\hline
Dataset & Duration (sec.) & Added triples (\%) & Deleted triples (\%)  & throughput (triples/sec.)\\
\hline
LUBM1K & 172.7 & 0 & 0 & 466,552\\
\hline
LUBM10K & 1,755.8 & 0 & 0 & 307,139\\
\hline
DBPedia & 29.2 & 0 & 0 & 478,129\\
\hline
Wikidata & 859.2  & 0 & 0.04 & 77,240\\
\hline
\end{tabular}
\caption{Ontology-based encoding with lite materialization}\label{liteMat}
\end{center}
\end{table*}

\begin{table*}
\begin{center}
\begin{tabular}{|c|r|r|r|}
\hline
Dataset & Duration (sec.) & Added triples (\%) & throughput (triples/sec.)\\
\hline
LUBM1K & 176.6 & 38.15 & 460,917\\
\hline
LUBM10K & 3,136.8 & 38.2 & 272,796\\
\hline
DBPedia & 57 & 13.6 & 356,348\\
\hline
Wikidata & 1,482 & 57.9 & 123,571\\
\hline
\end{tabular}
\caption{Ontology-based encoding with full materialization}\label{fullMat}
\end{center}
\end{table*}

\subsubsection{Query processing benefits}
We conduct experiments to quantify the benefits of LiteMat on querying large RDF databases.
We implement all the queries (detailed in Appendix~\ref{appendix:query}) in Scala using the translation method proposed in Sec.~\ref{sec:query}.
We execute queries $Q_1$ to $Q_4$ on  the LiteMat and on the full materialization of the LUBM10K dataset.
For baseline comparison, we also execute the rewritten queries $Q_1'$ to $Q_4'$ on the original LUBM10K dataset.
We first check for query completeness: each query consistently returns the same result set on the lite, full, and original dataset.
%
%
Then, we report on Table~\ref{tab:query} the response times of $Q_i$ queries for  LiteMat and or full materialization, and of $Q_i'$ queries for the no materialization database. 

\begin{table*}
\begin{center}
\begin{tabular}{|c|r|r|r|}
\hline
Query & Lite mat. & Full mat. &  No mat.\\
\hline
Q1 & 15.6 & 20.2 & 21.8\\
\hline
Q2 & 15.6 & 17.2 & 21.6\\
\hline
Q3 & 72.2 & 99.2 & 81.4\\
\hline
Q4 & 50.2 & 79.8 & 27.2\\
\hline
\end{tabular}
\caption{Query times (in sec.) for Lite/Full/No materialization}\label{tab:query}
\end{center}
\end{table*}

We observe that our solution (LiteMat) outperforms the baseline approach (no mat.) from 11\% (for $Q_3$) up to 28\% (for $Q_1$). This benefit is rather impressive considering that we implement the baseline approach in an optimized way using a conjunction of ``OR'' subqueries instead of a costly union of conjunctive subqueries.

On the opposite, we observe that query $Q_4$ containing 3 triple patterns, performs slower than its rewriting, because in this particular case, the rewriting consists of the union of an empty subquery (indeed, the original dataset does not contain any triple with the $Chair$ concept) and another simple subquery having only 2 triple patterns.
However, this is not a limitation of our approach since one can always simplify the query before executing it on the LiteMat database.

As expected, full materialization yields relatively poor performances because of the overhead implied by accessing a larger database.
The results show that LiteMat allows for fast query processing (thanks to the encoding) and the ability to store most of processed data in main-memory (a property due the Spark framework).

\section{Related works}
The first set of related works concerns semantic-aware RDF data encoding.
Most existing works in this field have been influenced by \cite{DBLP:conf/sigmod/AgrawalBJ89} which proposes to encode subsumption hierarchies with numeral values. This method proposes to encode the nodes of a tree in such a way that all the numerical values associated to subnodes of a supernode all range within an interval that can be computed from the supernode value. Although the work of \cite{DBLP:conf/sigmod/AgrawalBJ89} is essentially focusing on the hierarchies themselves, the approach was extended  in \cite{DBLP:conf/kr/Rodriguez-MuroC12}, with the so-called semantic index technique, to Ontology-Based Query Answering (OBQA) for the RDFS fragment of DL-Lite \cite{DBLP:journals/jar/CalvaneseGLLR07}. This later work is not providing any algorithmic details and does not consider the parallel computation of the dictionaries and of the dataset transformation. 
The work in WaterFowl \cite{DBLP:conf/esws/CureBRF14} also influenced the LiteMat project. Both systems are adopting a binary encoding of the elements of the TBox. 
In WaterFowl, this was mainly motivated by the use of Wavelet tree data structures whose nodes are composed of bit vectors, to store encoded RDF datasets. Rather than focusing on identifier intervals at  query answering time, WaterFowl uses a binary prefix encoding scheme to navigate in wavelet trees. Like the two other systems in this section, WaterFowl does not scale for large datasets due to the lack of a parallel computing approach.

Another set of related works concerns the distributed encoding of RDF datasets. The most recognized systems in this field are either using a dedicated computing environment \cite{DBLP:conf/esws/GoodmanJMAAH11}, the Map Reduce model \cite{DBLP:journals/concurrency/UrbaniMDSB13} or a parallel language running over a shared nothing architecture \cite{DBLP:conf/ccgrid/ChengKWT14}.
The approach presented in \cite{DBLP:conf/esws/GoodmanJMAAH11} uses a shared memory Cray XMT multi-million dollars machine to perform parallel hashing based on the linear probing scheme. The evaluation emphasizes that the method is efficient due to main-memory storage and performances are linear with the number of used cores. 
Nevertheless, the algorithm is specific to the shared memory architecture of the machine. 
The work of \cite{DBLP:journals/concurrency/UrbaniMDSB13} computes RDF encodings on top of the Hadoop framework using a standard cluster of commodity hardware (which was it the case of the aforementioned system).
The approach uses a sampling approach to encode the most popular terms which are cached in main memory. The encoding is performed through a set of map and reduce functions which can imply costly network communications. 
Moreover, the use of Hadoop's MapReduce implies I/O contention due to the intensive use of disk accesses. 
Finally, \cite{DBLP:conf/ccgrid/ChengKWT14} proposes an implementation using the X10 parallel programming language consisting of 3 steps: fragmentation of the data in equal-size chunks (based on terms hash value), encoding of these entries of these chunks are computed locally on each node of the cluster, much like our ABox encoding.
Note that none of these distributed encoding approaches make a distinction between the TBox and the ABox. Hence, their semantic-unaware encoding is not able to support inference optimization.
Moreover, our Spark-based implementation has the advantages of being easier to develop for and maintain due to its user-friendly, rich set of APIs. 

\section{Conclusion}
 In this paper, we have presented LiteMat, a data encoding scheme and query evaluation approach for the SPARQL query language. The main contribution of the system is an efficient and complete RDF query answering solution respecting the SPARQL's RDFS entailment regime. By taking the most of our clever TBox encoding, we are able to reduce the amount of materialized data and to limit the effort of query reformulation. Conducted evaluations emphasized the efficiency of the approach in terms of date storage size as well as encoding and query processing performances. Moreover, except for the TBox encoding, all processing steps of our system are implemented over the parallel computing framework Apache Spark, and thus produce high triple statements per second processing rates.

We consider that there is plenty of room for improvement and optimization for our system. For instance, some future works will concentrate on query optimization using some heuristics and statistics which can be computed during the ABox encoding time. We are also planning to support a more expressive entailment regime at query run-time, e.g., RDFS+ \cite{DBLP:books/daglib/0028543} which provides to specify properties characteristics (inverse, symmetric and transitive) and equality between individuals, classes and properties.

\section*{Acknowledgment}
This work has been partially supported by the ARESOS project from CNRS Program MASTODONS.

\bibliographystyle{abbrv}
\bibliography{amw15} 

\appendix
\section{Queries}Following are the SPARQL queries executed over both the LUBM1K and LUBM10K of our experimentation.
\label{appendix:query}
\subsection{Query $Q_1$: Inferences over the concept hierarchy}
\begin{verbatim}
SELECT ?x WHERE { ?x rdf:type Professor.}
\end{verbatim}
and its rewriting $Q_1'$:
\begin{verbatim}
SELECT ?x WHERE {{ ?x rdf:type Professor.} 
 UNION { ?x rdf:type AssistantProfessor.} 
 UNION { ?x rdf:type AssociateProfessor.} 
 UNION { ?x rdf:type Chair.}
 UNION { ?x rdf:type Dean.}
 UNION { ?x rdf:type Faculty.}
 UNION { ?x rdf:type VisitingProfessor.}
 UNION { ?x rdf:type FullProfessor.}}
\end{verbatim}
\subsection{Query $Q_2$: Inferences over the property hierarchy}
\begin{verbatim}
SELECT ?x ?y WHERE { ?x memberOf ?y.}
\end{verbatim}
and its rewriting $Q_2'$
\begin{verbatim}
SELECT ?x ?y WHERE {{?x memberOf ?y} UNION 
 {?x worksFor ?y} UNION {?x headOf ?y}}
\end{verbatim}
\subsection{Query $Q_3$: Inferences over the concept and property hierarchies}
\begin{verbatim}
SELECT ?x ?y WHERE { ?x rdf:type Professor.
 ?x memberOf ?y.}
\end{verbatim}
The rewriting, denoted $Q_3'$, of this query corresponds to the union of the cartesian product of the WHERE clauses of queries $Q_1$ and $Q_2$.
The query thus contains the union of 18 filters. For space limitation, we only display a subset of them: 
\begin{verbatim}
SELECT ?x WHERE {{?x rdf:type Professor.
 ?x memberOf ?y.} UNION {?x rdf:type 
 Professor. ?x worksFor ?y.} UNION {?x 
 rdf:type Professor. ?x headOf ?y.} UNION 
 {?x rdf:type AssistantProfessor. 
 ?x memberOf ?y.}...}
\end{verbatim}

\subsection{Query $Q_4$: Inferences over the property hierarchy}
This query is a generalized pattern of the query $Q_{12}$ from the LUBM benchmark. 
\begin{verbatim}
SELECT ?x WHERE { ?x rdf:type Chair.
 ?y rdf:type Department. ?x worksFor ?y.}
\end{verbatim}
and its rewriting $Q_4'$:
\begin{verbatim}
SELECT ?x WHERE {{ ?x rdf:type Chair.
 ?y rdf:type Department. ?x worksFor ?y.}
 UNION {?x headOf ?y. 
 ?y rdf:type Department.}} 
\end{verbatim}
\end{document}